# Contribution of Conceptual Modeling to Enhancing Historians' Intuition - Application to Prosopography

Jacky Akoka[1,2], Isabelle Comyn-Wattiau[3], Stéphane Lamassé[4], Cédric du Mouza[1]

[1] CEDRIC-CNAM, Paris, France, [2] Institut Mines Telecom-Business School, Evry, France
[3] ESSEC Business School, Cergy, France, [4] LAMOP-Sorbonne University, Paris, France

`jacky.akoka@lecnam.net,wattiau@essec.edu, stephane.lamasse@univ-paris1.fr,dumouza@cnam.fr`

**Abstract**. Historians, and in particular researchers in prosopography, focus a lot of effort on extracting and coding information from historical sources to build databases. To deal with this situation, they rely in some cases on their intuition. One important issue is to provide these researchers with the information extracted from the sources in a sufficiently structured form to allow the databases to be queried and to verify, and possibly, to validate hypotheses. The research in this paper attempts to take up the challenge of helping historians capturing and assessing information throughout automatic processes. The issue emerges when too many sources of uncertain information are available. Based on the high-level information fusion approach, we propose a process that automatically supports historians' intuition in the domain of prosopography. The contribution is threefold: a conceptual data model, a process model, and a set of rules combining the reliability of sources and the credibility of information.

## 1 Introduction

Historians, and in particular researchers in prosopography, generally study well documented small groups of elites and/or large groups of individuals, mostly anonymous and poorly documented. In both cases the reliability and the quality of the source material is crucial. Prosopography is a domain of research in the field of History related to the inquiry into the common characteristics of a group of historical actors by means of a collective study of their lives [Stone, 1971]. Prosopographical research has benefited from database technology, relying on relational databases to structure the data describing these individuals. These databases allow researchers to conjecture and test hypotheses. Hypothesis testing is a process that involves performing several steps ranging from the processing of the information contained in different and multiple sources to the validation of the hypothesis. This is not a simple transformation process. It gives rise to several interpretations due to the use of natural language. The latter is ambiguous, information is incomplete and imprecise. The structured representation of this information must also encode the uncertainty that characterizes it.

Data fusion researchers have proposed a process model that facilitates the integration of data and knowledge from several sources [Esteban *et al.*, 2005]. There are many analogies between information fusion and the process implemented by historians and researchers in prosopography, which captures information from historical sources to reconcile it, aiming at verifying historical hypotheses on the basis of a given population. We propose to adapt the information fusion process to the development of prosopographical databases, automatically gathering the information contained in the historical sources and aspiring the data of the existing databases, if any.

Our contribution is threefold: i) a conceptual data model gathering and representing the basic concepts of prosopography, ii) a conceptual process model adapting and enriching the high-level information fusion process, and iii) a set of rules ensuring the combination of the credibility of information and the reliability of sources to provide the researchers with automatically consolidated information. The combination of these three artifacts makes it possible to consolidate and further develop the intuition of researchers in History, and in particular of those in prosopography.

## 2      Related work

An important issue facing prosopographical researchers is how to make available the information extracted from different sources in a structured form to allow the databases to be queried and to verify, and possibly validate, hypotheses. Conceptual modeling helps to achieve this goal. Moreover, historians' ability to analyze this information by using their intuition may be limited by the fact that the information they capture originates from several sources and it can be inconsistent, ambiguous, and uncertain. Finally, it can be imprecise and incomplete. Providing the researchers in prosopography with an information fusion process will help them to develop a sharper intuition. The use of intuition is justified by the fact that History is very often seen as an interpretative discipline, allowing a plurality of opinions. In order to facilitate this interpretation, intuition is helpful, especially when expertise is not sufficient, when the problem and the data are unstructured, and finally when there is a lack of objective criteria [Barraclough, 1966].

### 2.1      Conceptual modeling for prosopography

Although there are several conceptual models to represent prosopographical data, the most common conceptual model is that of the factoid. Prosopography of Anglo-Saxon England (PASE) is an example of a factoid model [Bradley & Short, 2002]. Other examples include the Roman Republic project [Figueira & Vieira, 2017] and Charlemagne's European conceptual model[1]. [Akoka *et al.*, 2019] present a conceptual model that gathers and makes more generic the information contained in different prosopographical databases, namely the concepts of *People, Factoids, Places* and *Sources*. It also incorporates a representation of uncertainty.

---

[1] http://www.charlemagneseurope.ac.uk/

Note that not all prosopographical databases use the factoid conceptual model. For example, Studium Parisiense[2] uses a logical model without reference to the factoids. CHARM[3] is a reference model for cultural heritage. Conceptual modeling of prosopographical projects also calls on ontologies. The Factoid Prosopography Ontology (FPO), ontology of the PASE project, is based on OWL / RDFS. The CIDOC-CRM ontology is not based on factoids but on temporal entities which generalize the concept of event [Doerr, 2003]. It represents information relating to cultural heritage. To our knowledge, these ontologies do not take into account vagueness of information. An approach for the quantification of uncertain information can be found in [Martin-Rodilla *et al.*, 2019]. As it can be seen, there is not a single conceptual data model that encompasses the different viewpoints of prosopographical researchers.

## 2.2 Information fusion process

Information fusion is "an automated process which involves combining data in the broadest sense to estimate or predict the state of some aspects of a problem space of interest" [Snidaro *et al.*, 2016]. Information fusion, used to improve the quality of the information being considered, can greatly improve the imperfection that characterizes information captured by historians including information inconsistencies. Moreover, historians may have to make choices, but they have to face information conflicts that can be solved using information fusion approaches [Luo & Kay, 1988].

Many fusion approaches, models, and techniques have been proposed [Castanedo, 2013]. Data fusion approaches include probabilistic, evidential belief reasoning, and rough set-based fusion. Contributions in data fusion include [Alam *et al.*, 2018; Snidaro *et al.*, 2016; Pires *et al.*, 2016; Yao *et al.*, 2008; Menga *et al.*, 2020; Blasch *et al.*, 2012; Blasch *et al.*, 2013]. The objective of the first JDL Data Fusion model was to provide a process flow for sensor data fusion [Kessler *et al.*, 1991].

In summary, and to the best of our knowledge, there is not a prosopographical conceptual process model adapting and enriching the high-level information fusion process. Building on the analogy between the information fusion process and the manual process of the historian confronting different sources of information, we propose a conceptual data model and a process model that can be completely automated.

## 3 The conceptual data model

Information in one source can be supplemented, reinforced or contradicted by information in another source. When two pieces of information complement each other, in most cases, the incompleteness can be reduced. When two pieces of information are confirmed, the uncertainty is reduced. When they contradict each other, they can be juxtaposed and then be subject to the expertise of the historian who will relies on rules, on the credibility she grants to the source or on her interpretation of the context in which the author has written. A challenge is therefore to design a model which captures all this information and their cohabitation. .

---

[2] http://lamop-vs3.univ-paris1.fr/studium
[3] http://www.charminfo.org

### 3.1 The basic concepts of our approach

Our aim is to provide a unified model reconciling historians' viewpoints based mainly on four main concepts: *mention, factoid, fact,* and *hypothesis*.

**Definition 1.** *Mention: information found in a source.*
The mention is our finest grain information. It may be presented as affirmative (credibility=1) or with a level of uncertainty (0<credibility<1). It may also be defined with a given level of precision. An information fusion process for prosopographical research starts with source digitization, natural language processing (NLP) techniques encompassing the capture of mentions, followed by named entity recognition (NER) mechanisms to ensure the translation of mentions into factoids.

**Definition 2.** *Factoid: assertion based on one or several mentions with a certain level of agreement among historians.*
The factoid was coined in 1973 by [Mailer, 1973] as a fact which has no existence before appearing in a magazine or newspaper. In the factoid model, it denotes an assertion made by the project team that a source 'S' at location 'L' states something ('F') about person 'P'. A mention is an atomic factoid, in the sense that it cannot be decomposed into simpler elements. Moreover, it is linked to a unique source. It is also marked by uncertainty and imprecision.

**Definition 3.** *Fact: assertion that is known or proved to be true.*
Oppositely to a factoid, an assertion is a fact when it receives a large consensus to be true. A fact may result from the reconciliation of several similar factoids. As a consequence, it may be less precise, representing to some extent the greatest common divisor of several factoids. A fact achieves a broad consensus between historians, forming an historical truth. Some mentions may automatically be considered as facts given the status of the source where they come from.

**Definition 4.** *Hypothesis: tentative statement about the relationship between two or more facts and/or factoids.*
Research in humanities aims at proving or disproving hypotheses. The latter may not only be defined on facts but frequently imply less agreed assertions, represented as factoids in our context.

### 3.2 The conceptual data model

The conceptual model described in Fig. 1 is adapted from [Akoka *et al.*, 2019] in order to meet several requirements: i) encompass and put together the different viewpoints of prosopographical researchers; ii) conform to the information fusion process; iii) automatically propagate the certainty of information throughout the main concepts (mention, factoid, fact, hypothesis) as explained below (Fig. 2). A factoid may be composed of simpler factoids. If two sources describe the same information, they constitute two mentions in our model. The researcher interprets the different factoids leading to facts. The latter may be simple or may aggregate simple facts. Both factoids and facts, represented by the generic entity Datum, may support hypotheses. Factoids as well as facts are related to the other main concepts of proposopography, i.e. Persons, Places, and Time that we don't represent here for space reasons.

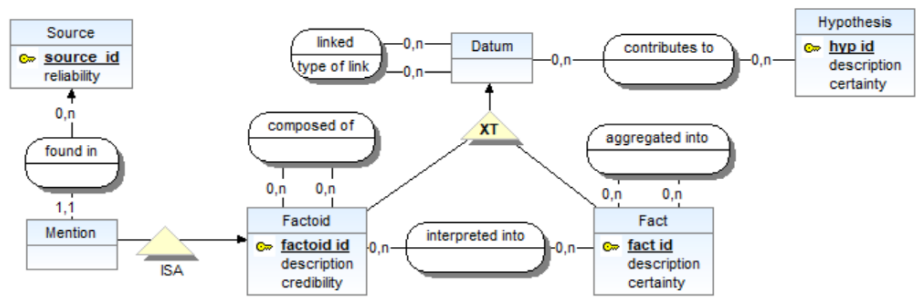

**Fig 1.** The conceptual data model (an excerpt)

This conceptual data model provides researchers with a model whose aim is to integrate the different factoid visions and to reconcile the different prosopographical approaches. It also supports our process model, the aim of which is to automate the building process of the historian's intuition. The combination of this data model and the process model presented below will save prosopography researchers' time, allow better handling of the uncertainty, and the possibility of going back to the source databases to interpret the uncertainty.

### 3.3 The conceptual process model

We describe the historical information fusion process by the means of a conceptual model (Fig. 2). To the best of our knowledge, it is the first time that the information fusion process model is adapted to digital humanities. The idea is to implement and make automatic the construction of the historian's intuition. Information capture (source preprocessing of information fusion) is based on NLP, ETL, and NER techniques and feeds the mentions of our conceptual data model. Credibility and reliability are also encoded. Making this step automatic releases the historian from this task so that she devotes herself to the generation of hypotheses. The certainty of the *mention* results from the combination (by a function to be defined) of the reliability of the source, combined with the credibility of the information as described in the source (for example, a very reliable source says: "it is very likely that…"). The main phases of our conceptual process model are described below.

**1. Information Capture phase**: basically, this phase consists in aspiring in *mentions*, information contained in the sources. We can assign a certainty combining the reliability of the source and the credibility of the information.

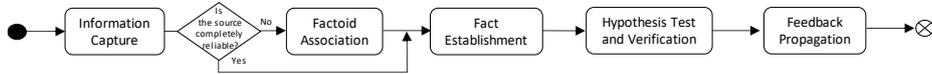

**Fig 2**. Process model

**2. Factoid Association phase** (level 1 of data fusion, or object refinement) applies association techniques (spatio-temporal alignment, correlation, clustering, grouping, etc.) to bring together statements that are similar and thus consolidate their certainty in a *factoid* of the same grain but with a higher certainty or construct a more structured *factoid* grouping together the different similar *mentions*. The "factoid association" combines all similar *factoids* to generate new richer or safer *factoids*. It compares the *factoids* based on their similar dimensions like the people involved for example. Thus, it can also spot people with similar names (using metrics like edit distance) involved in similar *factoids*. For example, if a new source of information is received, we can trigger the "factoid association" using forward chaining to deduce all the *factoids* resulting from this new arrival of information. In backward chaining, starting from a factoid, it searches for the simpler factoids that compose it. Factoid association generates factoids from factoids. We could formalize the concepts of the greatest common divisor and the least common multiple or max and min in a lattice.

**3. Fact Establishment phase** (level 2 of data fusion or situation assessment) makes it possible to combine factoids to converge on a fact. It combines different statistical techniques (removal of false positives, definition of a certainty threshold, etc.) and makes it possible to deduce facts. Again, we can fire a forward chaining in a classic application of the whole process from the capture of new information to, in cascade, enrich the entire database. We can also, in backward chaining, verify a hypothesis and for this seek the presence of supporting facts and the presence of factoids sufficiently convergent to attest the fact. In addition, for all factoids reaching a certain degree of confidence, one can establish a fact. One can also directly integrate in facts *mentions* since the reliability of the source is very strong.

**4. Hypothesis Test and Verification phase**. It is analogous to the previous one. It relies on *factoids* and *facts* to confirm or refute a hypothesis.

**5. Feedback propagation phase**. To our knowledge, this phase is new in the information fusion process. When a *hypothesis* is confirmed (resp. infirmed), it makes it possible to back propagate in the various *factoids* and in the information relating to the sources, the enhanced (resp. diminished) reliability of these sources. It is also during this phase that we can imagine thus downgrading a *fact*.

## 4    Illustrative example

Let us consider two different prosopographical data sources relating to historical figure Thomas Aquinas, considered as one of the main masters of scholastic philosophy and Catholic theology. Each source can be represented as an Uncertain Knowledge Graph (Fig.3), logical representation of our conceptual data model. For space reason, we cannot describe the mapping between conceptual and logical levels. Uncertain Knowledge Graph (UKG) have been recently investigated [Chen *et al.*, 2019]. They represent knowledge as a set of relations R defined over a set of entities E. More precisely, it consists of a set of weighted triples UKG={(t,s)}, with t=(e1,r,e2) a triple representing a relation fact where (e1,e2) ∈ $E^2$ and r ∈ R, and s ∈ [0,1] represents the confidence score for this relation fact to be true.

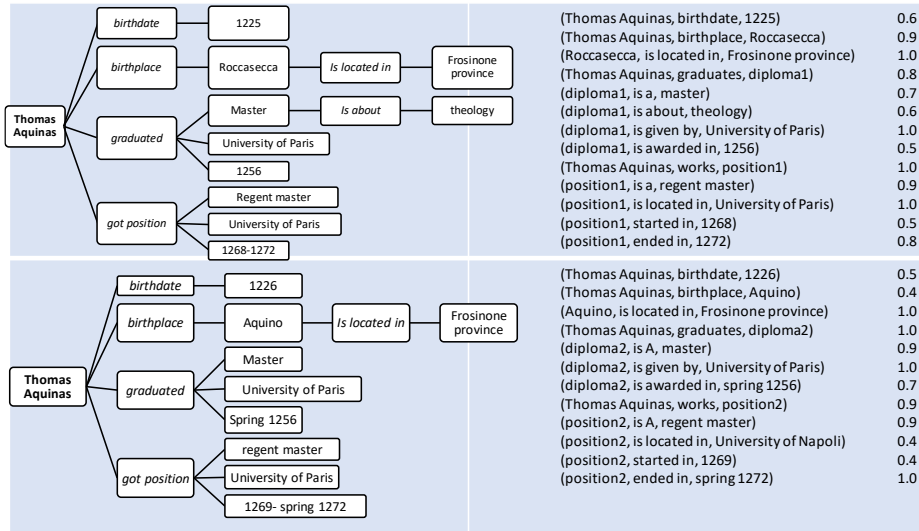

**Fig. 3**. Two prosopographical data sources represented as Uncertain Knowledge Graphs

A prosopographical data source can be considered as a collection of mentions. Basically a mention is a set of triples (subject, predicate, value) obtained using the n-ary relations transformation proposed by W3C[4]. Moreover, each triple is associated with a certainty score which represents the confidence we have for this relation to be true. As for UKG, we perform inference, for instance based on PSL logic [Chen *et al.*, 2019]. Since our objective is to integrate new prosopographical data, we must additionally define rules to combine information whose certainty can be reinforced or weakened. These rules are essentially expert rules. Thanks to these rules we can automate the process depicted in Fig. 2. We provide in the following some examples of these rules.

**Similarity detection:** Before applying a composition rule for two triples (s1, p, v1) and (s2, p, v2) we check whether s1 and s2 refer to the same subject. So we assume the existence of a similarity function for each domain, denoted $sim_D : D \times D \rightarrow \aleph$. This similarity is based on expert rules and largely differs from one domain to another.

---

[4] https://www.w3.org/TR/swbp-n-aryRelations/

**Composition rules:** Let T be the set of domains organized as tree (or upper semi-lattice), thanks to a predicate: is located in, is a, etc. For a domain D ∈ T and its associated predicate p, we denote $LCM_p : D \times D \rightarrow D$ the least common ancestor function for predicate. We also define the concept distance for a domain D ∈ T $dist_p(v1,v2)=min(dist_a(v1,LCM(v1,v2)), dist_a(v2,LCM(v1,v2)))$ where $dist_a(x,y)$ returns the difference between the level of y and x in the tree. So the closer one of the two values is to their LCM the smaller the concept distance is.

**Rule 1 [value-consistent predicate]:** Assume a predicate p, and codomain(p) ∈ T, we have: $\{(s, p, v1), p1\} \wedge \{(s, p, v2), p2\} \wedge dist_p(v1,v2) \leq \tau$
$$\rightarrow \{(s, p, LCM(v1,v2)), aggreg\_lcm(p1,p2)\}$$

Aggreg_lcm is an aggregation function for the two probabilities. For consistent values it can be for instance max, avg, min, or 1-(1-p1) x (1-p2). τ is the distance threshold for the concept distance. Its objective is to avoid to deduce some information with an important loss of precision (for instance LCM(Paris,Versailles)=Parisian Region is accepted, but LCM(Paris, Roma)=Europe is not).

**Rule 2 [value-inconsistent predicate]** Assume a predicate p, and codomain(p) ∈ T. we have: $\{(s, p, v1), p1\} \wedge \{(s, p, v2), p2\} \wedge dist_p(v1,v2) \wedge p1 > p2 > \tau$
$$\rightarrow \{(s, p, v1), aggreg\_inconsistent(p1,p2)\}$$

Aggreg_inconsistent is an aggregation function for the two probabilities. For inconsistent values it can be for instance min(p1, p2), p1-p2, or p1x(1-p2)

**Rule 3 [Fact building]:** Consider the set of all triples associated to a subject s, denoted Ω, and a threshold value π set for the system, that we call the fact threshold. We also denote with $Ω^+$ the subset of Ω composed by the triples with a certainty value greater than π. Then we decide that $Ω^+$ is a fact while Ω is a complex factoid.

*In the example, if π = 0.9, 'Thomas Aquinas got a diploma and this diploma is a master' is a fact. Thomas Aquinas got a diploma and this diploma is a master in 1256 is a factoid. Assume we have a source S3: {(Thomas Aquinas, graduates, diploma3), 1.0} and {(diploma3, is A, doctorate), 0.4} and {(diploma3, is awarded in, 1256), 0.9}*

*→: {(Thomas Aquinas, graduates, diploma2), 0.99} and {(diploma2, is A, master), 0.58} and {(diploma2, is awarded in, 1256), 0.98}. So now, the fact is Thomas Aquinas got a diploma and this diploma was received in 1256 and no more the previous one.*

In the previous example, we tried to illustrate how our conceptual model may be derived into a logical knowledge graph and how our information fusion process may propagate the credibility of the mentions extracted from the sources to the facts but also infirm facts.

## 5 Validation of the approach

[Prat *et al.*, 2015] consider that a design science approach generally results in a set of artifacts constituting a system and proposes to validate the properties of this system. We use their framework to present the different validations we conducted. In italics, we mention the criteria that we were able to check.

The system composed of these two conceptual models is goal-directed. We mainly tried to assess the *goal attainment* criterion. The criterion *added value* is twofold. The combination of factoids assigned with credibility improves the state-of-the-art by providing the historians with an enriched set of information that would require too much time to be explored directly. Their confidence is based on the fact that they initially define the reliability of the sources, which is one component of the credibility computed on factoids. The second added value is due to the feedback propagation step which automatically updates the reliability of sources resulting from their confrontation.

In order to check the *utility* of the approach, we organized a focus group with historians. We explained the data model and the process model and questioned them on the ability of the approach to represent their reasoning. They were able to validate our objective: better define a process that can then be automated and considerably improve the state-of-the-art by embedding the codification of the credibility. They could also enrich our understanding by providing us with several publications where historians tried to explain their reasoning.

We also questioned the historians to check the *completeness* and the *minimality* of our conceptual data model. In our first version, the **mention** was an entity related to **factoid** but not a subtype of it. This idea was first considered as interesting. But they were unable to clearly distinguish between **mentions** and **factoids** whereas they were immediately convinced by the interest of representing both **factoids** and **facts**.

Regarding the conceptual process model, they think that it really represents their reasoning when analyzing directly information sources. In particular, depending on the category of sources (e.g. registers of universities), one can immediately translate the information contained in terms of **facts**. In other sources (e.g. testimonies), usually only **factoids** can be deduced and a credibility may be associated. In this sense, we can consider that we checked the *fidelity* of the process model *to the phenomenon it represents* (the construction of the historian's intuition).

This validation allowed us to check its *technical feasibility* using the mapping to knowledge graphs and testing some inference rules.

## 6   Conclusion and future research

The research question discussed in this paper is how to enhance historians' intuition. We answered the research question by developing an approach based on the combination of a conceptual data model and a conceptual process model. The conceptual data model captures the main concepts of prosopography while the process model implements and makes automatic the construction of the historian's intuition. This model is enriched with a feedback propagation step, useful when a hypothesis tested by historians is confirmed (resp. infirmed), allowing them to back propagate the results in the information relating to the historical sources. We also provided a set of rules combining the reliability of sources and the credibility of information manipulated by historians, in particular in prosopography. We illustrated the approach using two different data sources, represented as uncertain knowledge graphs. Finally, an argumentation is presented to validate our approach. Future research will include more experimentations and a comprehensive implementation, leading to the proposition of additional rules, and alternative methods for combining credibility depending on the context.